\documentclass[paper,a4paper]{JHEP3} 

\JHEPspecialurl{http://jhep.sissa.it/JOURNAL/JHEP3.tar.gz}

\usepackage{epsfig,multicol,bbm}
\usepackage{amssymb}
\usepackage{amsmath}
\usepackage{cite}

\newcommand\fverb{\setbox\pippobox=\hbox\bgroup\verb}
\newcommand\fverbdo{\egroup\medskip\noindent%
                        \fbox{\unhbox\pippobox}\ }
\newcommand\fverbit{\egroup\item[\fbox{\unhbox\pippobox}]}
\newbox\pippobox
\newcommand{\beq}{\begin{equation}}
\newcommand{\eeq}{\end{equation}}
\newcommand{\bea}{\begin{eqnarray}}
\newcommand{\eea}{\end{eqnarray}}
\newcommand{\bem}{\begin{multline}}
\newcommand{\eem}{\end{multline}}
\newcommand{\beg}{\begin{gather}}
\newcommand{\eeg}{\end{gather}}
\newcommand{\uk}{\underline{k}}

\newcommand{\noi}{\noindent}
\newcommand{\nn}{\nonumber}
\newcommand{\dgg}{^{\dagger}}

\newcommand{\gt}{>}
\def\parf{\stackrel{\leftrightarrow}{\partial_i}}
\def\tr{\mbox{tr}\{}
\def\t{t}
\def\d{\delta}
\def\eq#1{{Eq.~(\ref{#1})}}
\def\fig#1{{Fig.~\ref{#1}}}

\title{${\cal O}(\alpha_s^2)$-corrections to JIMWLK evolution from the
classical equations of motion}

\author{Javier L. Albacete\thanks{Supported by the U.S. Department of Energy
 under Grant No. DE-FG02-05ER41377.}~
                \\
        Department of Physics, The Ohio State University,
                \\
        191 W. Woodruff Avenue, 43210 Columbus, OH, USA
                \\
        E-mail: \email{albacete@mps.ohio-state.edu}}

\author{N\'estor Armesto\thanks{Supported by Ministerio de Educaci\'on
y Ciencia of Spain under a contract Ram\'on y Cajal, and by
CICYT of Spain under projects FPA2002-01161 and FPA2005-01963.}~
                \\
        Departamento de F\'{\i}sica de Part\'{\i}culas and
IGFAE,
Universidade de Santiago de Compostela,
 15782 Santiago de Compostela, Spain\\
        E-mail: \email{nestor@fpaxp1.usc.es}}

\author{Jos\'e Guilherme Milhano\thanks{Partially supported by
the Funda\c
c\~ao para
a Ci\^encia e a Tecnologia of Portugal under contract
SFRH/BPD/12112/2003.}~
                \\
CENTRA, Instituto Superior T\'ecnico (IST),
Av. Rovisco Pais, P-1049-001 Lisboa, Portugal,
and\\
Departamento de F\'{\i}sica, FCT, Universidade do Algarve,
P-8000-117 Faro, Portugal\\
        E-mail: \email{gui@fisica.ist.utl.pt}}

\abstract{In this work we calculate some
$\mathcal{O}(\alpha_s^2)$ corrections to 
the JIMWLK kernel in the framework of the light-cone wave function approach to
the high energy limit of QCD. The contributions that we consider originate 
from higher order corrections in the strong coupling and in the 
density of the projectile to the solution of the classical equations of motion
that
determine
the Weizs\"acker-Williams fields of the projectile.
We study the structure of these corrections in the dipole limit, showing 
that they are subleading in the large-$N$ limit and that they cannot be fully
recast in the form of dipole degrees of freedom, but rather present a
complicated color structure. }

\begin{document} 



\section{Introduction}
\label{intro}

The contemporary description of the  high energy evolution of QCD scattering
amplitudes, given by the B-JIMWLK equations, has been developed over the last
decade
\cite{McLerran:1993ni,McLerran:1993ka,McLerran:1994vd,Jalilian-Marian:1996xn,Jalilian-Marian:1997jx,Jalilian-Marian:1997gr,Jalilian-Marian:1997dw,Jalilian-Marian:1998cb,Kovner:1999bj,Kovner:2000pt,Weigert:2000gi,Iancu:2000hn,Iancu:2001ad,Iancu:2001md,Ferreiro:2001qy,Balitsky:1995ub,Balitsky:1997mk,Balitsky:1998ya,Mueller:2001uk,Blaizot:2002xy}
building   
upon the pioneering ideas on gluon saturation set out by Gribov, Levin and
Ryskin \cite{Gribov:1984tu} in the early 80s. Notwithstanding that the
B-JIMWLK set of equations has proved difficult to tackle and 
that its complete solution remains unknown, a series of numerical
\cite{Braun:2000wr,Kimber:2001nm,Armesto:2001fa,Levin:2001et,Lublinsky:2001bc,Albacete:2003iq,Albacete:2004gw,Golec-Biernat:2003ym,Rummukainen:2003ns} 
and analytical
\cite{Iancu:2002tr,Mueller:2002zm,Mueller:2003bz,Munier:2003vc,Munier:2004xu,Munier:2003sj}
studies have established the asymptotic properties of the B-JIMWLK equations
by considering their mean field limit, where the  B-JIMWLK set reduces to a
single closed equation --- the Balitsky-Kovchegov (BK) equation
\cite{Balitsky:1995ub,Balitsky:1997mk,Balitsky:1998ya,Kovchegov:1999yj,Kovchegov:1999ua}. 
Further, the results from this mean field equation deviate at most 10\% from
those obtained numerically from the full B-JIMWLK \cite{Rummukainen:2003ns}.

Significantly, the  B-JIMWLK scheme neglects important effects
\cite{Levin:2003nc,Mueller:2004se,Iancu:2004es,Iancu:2004iy} ---
variably referred to as ``Pomeron loops''
\cite{Iancu:2004iy,Mueller:2005ut,Levin:2005au,Iancu:2005nj,Marquet:2005hu,Kovner:2005en,Kovner:2005uw,Kovner:2005nq}, 
``fluctuations'' \cite{Iancu:2004es} or ``wave 
function saturation effects''  --- due to gluon fluctuations.
The insufficiencies of the B-JIMWLK framework become apparent once one recalls
that these equations were derived under the explicit assumption --- and,
therefore, are strictly valid for such a physical situation --- of large
target gluon density and of a dilute projectile. 
Therefore, the B-JIMWLK formalism fails to properly describe the high-energy
evolution for less asymmetrical systems (i.e. those where both target and
projectile are dense). These crucial observations have led to a spurt of
activity in this domain targeted at obtaining an evolution scheme correctly
accounting for gluon fluctuation effects 
\cite{Mueller:2004se,Iancu:2004iy,Mueller:2005ut,Levin:2005au,Iancu:2005nj,Marquet:2005hu,Kovner:2005en,Kovner:2005uw,Kovner:2005nq,Kovner:2005jc,Soyez:2005ha,Enberg:2005cb,Marquet:2005ak,Hatta:2005rn,Blaizot:2005vf,Kozlov:2006cg,Armesto:2006ee,Marquet:2006xm}.
A duality transformation linking the low and high density regimes
\cite{Kovner:2005en,Kovner:2005uw} places strict 
constraints on the form of any would-be complete evolution kernel.

Different paths have been followed in attempting to write an evolution kernel
including dynamics beyond B-JIMWLK. A potentially fruitful path explores the
connection between high energy QCD evolution and a reaction-diffusion process, 
\cite{Iancu:2004es} to suggest that the full dynamics should be described by
an equation belonging to the universality class of the stochastic
Fisher-Kolmogorov-Petrovsky-Piscunov (sFKPP) equation \cite{Iancu:2004iy,Soyez:2005ha,Enberg:2005cb,Marquet:2005ak,Iancu:2005nj,Brunet:2006zn}. A different
strategy, the light-cone wave function approach  
\cite{Kovner:2001vi,Kovner:2005en,Kovner:2005uw,Kovner:2005nq,Kovner:2005jc},
yields an evolution kernel not subject to the restrictions underlying the
B-JIMWLK equations. This kernel, written in terms of the classical gluon
fields of the projectile, reduces to the JIMWLK kernel when the fields are
taken at leading order in the charge density. In this work we go one step
further by computing the leading projectile density corrections.

This paper is organized as follows: in Sec.~\ref{setup} we briefly introduce
the light-cone wave function approach to high energy QCD.
In
Sec.~\ref{results} we present our general results for the leading projectile
density correction, and in Sec.~\ref{dipole} we thoroughly discuss the dipole
limit of the results of the previous section. Our conclusions are presented in
Sec.~\ref{conclu}.

\section{Setup}
\label{setup}

We begin by reviewing the main features of the light-cone wave function
approach to the high energy limit of QCD as derived by Kovner, Lublinsky and
Wiedemann in \cite{Kovner:2005jc,Kovner:2001vi}.
Consider the collision between a bunch of energetic gluons moving in the
light-cone `+' direction --- the projectile ---  with a dense hadronic target 
characterized by large gluon fields. Let $Y$ be the total rapidity of the
collision. The light-cone wave function of the 
incoming projectile is given by
\beq
|\Psi^{init}(Y)\rangle = \Psi[a_i^{a\dagger}(k^+,\uk)]|0\rangle,
\label{psii}
\eeq
where $a^{a\dagger}_i(k^+,\uk)$ are the creation operators for gluons with
color index $a$,
longitudinal momentum above some hard cutoff, $k^+\!\!\gt\!\!\Lambda^+$, and 
transverse momentum $\uk$. Assuming that each of the gluons in the projectile
interacts independently with the target, the wave function after the
interaction is given by 
\beq
|\Psi^f(Y)\rangle = \Psi[S^{ab}(x_i)a_i^{b\dagger}]|0\rangle,
\label{psif}
\eeq
where $S^{ab}(x_i)$ is the $S$-matrix corresponding to the propagation
through the target of a single gluon located at transverse
position $x_i$. 
At very high
energies the interaction with the target eikonalizes and the $S$-matrix is
diagonal in the transverse coordinates of the incoming gluons. Moreover, the 
$S$-matrix elements depend only on the gluon fields in the target, $A_t^{\mu}$,
which allows us to take them,  rather than the target fields themselves, as the physical degrees of freedom to describe the target.
Although
$S^{ab}$ is a quantum operator acting on the Hilbert space of the target
fields, the commutators of $S$-operators are suppressed by powers
of the target density, which hereinafter is assumed to be very large. 
Therefore $S^{ab}$ can, and will, be considered as a classical $c$-number, 
in the spirit of the semiclassical approach to dense gluonic systems \cite{McLerran:1993ni,McLerran:1993ka,McLerran:1994vd}.
With all this in mind, the scattering matrix of the projectile at
rapidity $Y$ can be written as
\beq
\Sigma_Y[S]=\langle \Psi^i(Y)|\Psi^f(Y)\rangle.
\label{sigmai}
\eeq
High energy processes are characterized by a large separation of time
scales --- i.e. the time needed for the fast projectile to propagate through the
target (the interaction time)  is much shorter than the typical 
time scale under which the large, soft target fields vary. 
Thus, the fast projectile probes the target fields in a fixed, frozen 
configuration. Consequently, the physical scattering matrix is obtained by 
averaging over all the possible configurations of the gluon fields in the 
target:
\beq
\langle \Sigma \rangle = \int dS \,\Sigma[S]\,W_Y[S],
\label{psigma}
\eeq
where the weight functional $W_Y[S]$ can be understood as the probability
density  for the target to
have a certain configuration of the fields, $S$, at rapidity $Y$ (see \cite{Kovner:2005nq} 
for a detailed discussion of the physical meaning and properties 
of $W_Y[S]$). 
In the wave function approach, the energy evolution of the system described 
above is achieved by boosting the projectile to higher rapidities, leaving 
the target unevolved. In this 
way all the information about the energy evolution of the system is encoded 
in the behavior of the projectile wave function as
opposed to the strategy followed in the original derivation of the JIMWLK 
equation, in which the quantum fluctuations originated from boosting the target
to higher energies were resummed in the presence of 
strong background fields, leading to the renormalization of the weight
functional, $W_Y$. 

To first order in $\d Y$ the projectile wave function at rapidity $Y+\d Y$ 
is given by:
\bea
|\Psi^i(Y+\delta Y)\rangle &=& \left\{1-\frac{1}{2}\delta Y\int
d^2z\,b_i^a(z,[\rho])b_i^a(z,[\rho])\right.\nonumber\\
 &+&\left. i\int d^2z\,b_i^a(z,[\rho])\int_{(1-\delta
Y)\Lambda}^{\Lambda}\frac{dk^+}{\sqrt{\pi}|k^+|^{1/2}}\,a_i^{\dgg
a}(k^+,z)\right\}|\Psi^i(Y)\rangle,
\label{evpsii}
\eea
where $b_i$ are the Weizs\"acker-Williams (WW) fields of the projectile, which
depend uniquely on the projectile density operator $\rho$, defined as:
\beq
\rho^a=\,\int_{\Lambda^+}\!dk^+a_i^{\dagger b}(k^+,z)T_{bc}^a\,a_i^c(k^+,z),
\label{rho}
\eeq
where $T^a$ are the generators of $SU(N)$ in the adjoint representation.
For a dilute projectile like the one considered here, the number of gluons in
its wave function is small $\rho\sim\mathcal{O}(1)$.
The physical meaning of \eq{evpsii} is clear: The hard, `valence' gluons in the
initial wave function are dressed with a cloud of soft gluons, the
Weizs\"acker-Williams (WW) fields $b_i$. These fields are determined from the 
classical Yang-Mills equations of motion (EOM),
in which the hard gluons enter as an external
source. The separation between soft and hard modes is made at an arbitrary 
scale $\Lambda^+$. The boost of the projectile opens up the phase space 
for the production of new hard gluons out of the soft WW fields. 
This production process is accounted for by the
last term in the rhs of \eq{evpsii}. The first term corresponds to no 
gluon production, whereas the second term corresponds to virtual corrections
required to ensure the right normalization of the wave function.  

The wave function of the
projectile after the collision with the target is given by an analogous
expression:
\bea
|\Psi^f(Y+\delta Y)\rangle &=& \left\{1-\frac{1}{2}\delta Y\int
d^2z\,b_i^a(z,[S\rho])b_i^a(z,[S\rho])\right.\\
 &+ & \left. i\int d^2z\,b_i^a(z,[S\rho])\int_{(1-\delta
Y)\Lambda}^{\Lambda}\frac{dk^+}{\sqrt{\pi}|k^+|^{1/2}}S^{ab}(z)a_i^{\dgg
b}(k^+,z)\right\}|\Psi^f(Y)\rangle,\nonumber
\label{evpsif}
\eea
where now the WW fields, $b_i$, are given by the solution of the classical EOM 
for rotated sources $S^{ab}\rho^b$. The scattering matrix of the evolved 
system is:
\beq
\Sigma[S](Y+\delta Y)=\langle \Psi^i(Y+\delta Y)|\Psi^f(Y+\delta Y)\rangle.
\label{sigmaf}
\eeq
From \eq{sigmai} and \eq{sigmaf}, it is straightforward to derive an
evolution equation for the scattering matrix $\Sigma[S]$ which, thanks to the
Lorentz invariance of \eq{psigma}, can
be converted into an evolution equation for the target weight functional,
$W_Y$:  

\beq
{\delta W[S]\over \delta Y}=\chi\, W[S], 
\label{evw}
\eeq
with the kernel of the evolution given by

\beq
\chi=-{1\over 2\pi}\int_z\,[b_i^a(z,[\rho])b_i^a(z,[\rho])+
b_i^a(z,[S\rho])b_i^a(z,[S\rho])-2b_i^b(z,[\rho])b_i^a(z,[S\rho])
S^{ba}(z)].
\label{kerb}
\eeq

Importantly, \eq{kerb} reduces to the JIMWLK kernel when the classical
fields are taken at leading order in the charge density of
the projectile, $g\rho$. Our goal is to derive higher order corrections 
to the JIMWLK evolution by solving the classical 
equations of motion at next order in $g\rho$.

At this point it should be noted that the expressions in \eq{evpsii} and 
\eq{evpsif} for the evolved wave function are not complete, and they are
correct only in the limit of small projectile density. The more general 
expression for the evolved wave function as given by Eq. (2.5) in 
\cite{Kovner:2005nq} includes an extra Bogolyubov 
transformation with respect to the expression in Eqs.
(\ref{evpsii}) and (\ref{evpsif}) 
in this paper. It is argued in \cite{Kovner:2005nq} and assumed in this work
that such transformation 
reduces to the unity operator in the limit of small projectile density,
$\rho\rightarrow 0$. 
Henceforth we restrict ourselves to the study 
of high density corrections to the kernel of the evolution arising from the 
expansion of the classical gluon fields, $b_i$, in terms of 
the projectile charge density, $g\rho$.

It is convenient to rewrite the kernel of the evolution 
in terms of left and right rotation operators, whose action on the scattering
matrix of a gluonic projectile is defined as:
\bea
J_L^a(x)\,\Sigma[S]&=&-\mbox{tr}\left\{T^aS(x)\frac{\d}{\d
    S^{\dagger}(x)}\right\}\,\Sigma[S]\, ,\\
J_R^a(x)\,\Sigma[S]&=&-\mbox{tr}\left\{S(x)T^a\frac{\d}{\d
    S^{\dagger}(x)}\right\}\,\Sigma[S]\, .
\eea
In terms of these operators, the kernel of the evolution reads
\beq
\chi=-\frac{1}{2\pi}\int_z\,[b_i^a(z,J_L)b_i^a(z,J_L)+
b_i^a(z,J_R)b_i^a(z,J_R)-2b_i^b(z,[J_L])b_i^a(z,[J_R])
S^{ba}(z)].
\label{kerbj}
\eeq

\section{General results}
\label{results}
To determine the WW fields entering the kernel of the evolution, \eq{kerbj},
we need to solve the classical
Yang-Mills equations of motion in the light-cone gauge, $A^+\!=\!0$, 
with the fast 
valence gluons playing the role of an external current. 
Since the hard gluons are fast moving in the `+' direction, the current can be
written as $J^{\nu}(z)=g\,\delta^{\nu+}\delta(z^-)\rho(z)$, and the
equations read  
\bea
D_{\mu}F^{\mu\nu}(z)=J^{\nu}(z)\, ,\label{ym1} \\
D_{\mu}J^{\mu}(z)=0\, ,
\label{ym2}
\eea
where \eq{ym2} ensures the covariant conservation of the current. Since the
number of gluons in the projectile is assumed to be small,  
$\sim\mathcal{O}(1)$, the current that they generate is of order $g$, 
$J\sim\mathcal{O}(g)$. Thus, $g$ will be the small parameter that controls our
expansion of the solution.

The general solution of these equations has been extensively 
discussed in \cite{Kovchegov:1996ty,Kovchegov:1997pc}. In particular, it was shown in \cite{McLerran:1993ni,McLerran:1993ka,McLerran:1994vd} that it 
is consistent to look for 
`static' solutions of \eq{ym1} --- i.e, solutions independent of $x^+$, 
with $A^-=0$. Such a solution 
is a pure gauge with just transverse components. Under these assumptions
\eq{ym2} is trivially satisfied and, by writing the transverse components of the
field as $A_i(z^-,z)=\theta(z^-)b_i(z)$,  Eq. (\ref{ym1}) becomes 
\bea
\partial_ib^a_i(z)+gf^{abc}\,b_i^b(z)b_i^c(z)=g\rho^a(z)\, ,\\
\partial_ib^a_j(z)-\partial_jb^a_i(z)+gf^{abc}\,b_i^b(z)b_j^c(z)=0\, ,
\label{ym3}
\eea
Expanding the solutions in powers of $g$:
\beq
b^a_i(z)=g\left(b^a_{i1}(z)+g^2\,b^a_{i2}(z)+\mathcal{O}(g^4)\right),
\label{ex}
\eeq
we get
\bea
b^a_{i1}(z)=\frac{1}{2\pi}\int d^2x\frac{(z-x)_i}{(z-x)^2}\,\rho^a(z)=
\frac{1}{2\pi}\int \!d^2x\,\left(\partial_iX\right)\rho^a(z),\\
b^a_{i2}(z)=-\frac{1}{4(2\pi)^2}f^{abc}\!\int d^2x\,d^2y\,(X\!\parf
\!Y)\rho^b(x)\rho^c(y),
\label{bsol}
\eea
where we have made use of the following shorthand notation for the coordinate
dependence of the solution:
\bea
\partial_iX\equiv \partial^{z}_i \ln\left(|z-x|\lambda\right)\, , \nn \\
X\!\parf\! Y\equiv X\left(\partial^{z}_iY\right)
-\left(\partial^{z}_iX \right)Y\, ,
\label{not}
\eea
where $\partial^{z}_i$ denotes the partial derivative with respect to the transverse
components of $z$, $i=1,2$.

The expansion of the solution in \eq{ex} immediately translates into an 
expansion for the kernel of the evolution in powers of $\alpha_s=g^2/4\pi$: 
\beq
\chi=\alpha_s\left(\chi^{1)}+\alpha_s\,\chi^{2)}+\mathcal{O}(\alpha_s^2)
\right),
\label{expk} 
\eeq
whose leading term is the JIMWLK kernel: 
\bea
\alpha_s\chi^{1)}=\chi^{JIMWLK} &=& -\frac{\alpha_s}{2\pi^2}\int_{xyz}
\!\!(\partial_iX)(\partial_iY)\left[J_L^a(z)J_L^a(z)\right. \nn \\
 & &+\left.J_R^a(z)J_R^a(z)-2J_L^a(z)J_R^b(z)S^{ab}(z)\right].
\label{kjimwlk}
\eea
The first correction to JIMWLK, $\chi^{2)}$, is obtained by keeping 
terms up to order $\alpha_s^2$ in the product WW fields, yielding 

\bea
\chi^{2)}&=&-\frac{1}{(2\pi)^2}\int_{xywz}f^{abc}\,(\partial_i Y)
(X\!\parf\! W)\,
\left[J_L^a(x)J_L^b(y)J_L^c(w)+J_R^a(x)J_R^b(y)J_R^c(w)\right]\nn  \\
& &-2f^{acd}(\partial_i X)(Y\!\parf\! W)S^{ba}(z)\left[J_L^b(x)J_R^c(y)
J_R^d(w)+J_L^b(x)J_L^c(y)J_R^d(w)\right].
\label{ker}
\eea

Note that from the second order solution of the EOM we could immediately 
derive part of the $\mathcal{O}(\alpha_s^3)$ corrections to the kernel. 
However,  a complete derivation of the $\mathcal{O}(\alpha_s^3)$ term in 
\eq{expk} would require the $O(g^5)$ solution to the EOM, which is 
beyond the scope of this paper.

\section{Dipole model limit}
\label{dipole}
The color structure of a generic projectile composed of gluons can be rather
complicated, consisting of different color multipoles mix with 
each other through the evolution in a highly non-trivial way. However,  
in the large-$N$ limit this complicated
structure is greatly simplified and the high energy evolution can be recast
in
terms of dipole degrees of freedom. 
More precisely, the JIMWLK equation is equivalent to an infinite 
hierarchy of coupled differential equations for the
correlators of the gluon fields.  In the large-$N$ limit the whole hierarchy
decouples and one is left with a single, closed, non-linear evolution 
equation for the dipole scattering amplitude --- the BK equation \cite{Balitsky:1995ub,Balitsky:1997mk,Balitsky:1998ya,Kovchegov:1999yj,Kovchegov:1999ua}. 

In this section we explore the color structure of the correction to
JIMWLK evolution derived in the previous section.
In order to do so, we consider an initial projectile entirely
describable by dipole degrees of freedom:
\beq
\Sigma=\Sigma[s],
\label{Sdip1}
\eeq
where
\beq
s(x,y)=\frac{1}{N}\tr S_F\dgg(y)S_F(x)\}
\label{sdip}
\eeq
is the scattering matrix for a $q$-$\bar{q}$ dipole, with $x$ and $y$ the
transverse coordinates of, respectively,  the quark and  the antiquark. The
subscript $F$ in the rhs of
\eq{sdip} indicates that the scattering matrix is to be taken for
particles in the fundamental representation of $SU(N)$, and will be usually
omitted in
the following.  
Our purpose is
to study the resulting color structure of the projectile under evolution, i.e.
 we want to calculate $\chi^{2)}\,\Sigma[s]$. 
The action of the left and right rotation operators on a dipole-like
projectile scattering matrix is given by:
\bea
J_L^a(x)\,s[u,v]&=&-\frac{1}{2}\,\mbox{tr}\left\{t^aS(x)\frac{\d}{\d S^T(x)}-
S^{\dagger}(x)t^a\frac{\d}{\d S^*(x)}\right\}\,s[u,v]\, ,\label{jlf}\\
J_R^a(x)\,s[u,v]&=&-\frac{1}{2}\,\mbox{tr}\left\{S(x)t^a\frac{\d}{\d S^T(x)}-
t^aS^{\dagger}(x)\frac{\d}{\d S^*(x)}\right\}\,s[u,v]\, , \label{jrf}
\eea
where $t^a$ are now the $SU(N)$ generators in the fundamental representation.
From Eqs. (\ref{jlf}) and (\ref{jrf}) it can be proved that both
$J_L$ and $J_R$ are hermitian operators, a property which immediately 
ensures the hermiticity of the kernel of the evolution.
\subsection{General expressions}
\label{general}

Using the following notation for the functional derivatives of the scattering
matrix: 
\beq
\d_{uv;\dots;qt}\Sigma\equiv\left[\frac{\d}{\d s(u,v)}\cdots\frac{\d}{\d
    s(q,t)}\right]\,\Sigma[s], 
\label{otra}
\eeq
and defining $S_z\equiv S_F(z)$,
we can write the action of the different pieces of the kernel in \eq{ker} on
$\Sigma[s]$ as
\bea
\label{rrr}
f^{abc}&&\!\!\!\!J_R^a(x)J_R^b(y)J_R^c(w)\Sigma[s]= 
-\frac{1}{N}f^{abc}[\d(\!v\!-\!w\!)\!-\!\d(\!u\!-\!w\!)]\Bigl. 
\Bigr\{  \\
& &\qquad[\d(\!v\!-\!y\!)\!+\!\d(\!u\!-\!y\!)]
[\d(\!v\!-\!x\!)\!-\!\d(\!u\!-\!x\!)]
\tr S\dgg_uS_v\t^a\t^b\t^c\}\d_{uv}\Sigma \nn \\
& &+\frac{1}{N}[\d(\!v\!-\!y\!)\!+\!\d(\!u\!-\!y\!)]
[\d(\!r\!-\!x\!)\!-\!\d(\!p\!-\!x\!)]
\tr S\dgg_uS_v\t^b\t^c\}\tr S\dgg_pS_r\t^a\}\d_{pr;uv}\Sigma \nn\\
& & +\frac{1}{N}[\d(\!r\!-\!y\!)\!
-\!\d(\!p\!-\!y\!)][\d(\!v\!-\!x\!)\!+\!\d(\!u\!-\!x\!)]
\tr S\dgg_uS_v\t^a\t^c\} \tr S\dgg_pS_r\t^b\}\d_{pr;uv}\Sigma \nn \\
& & +\frac{1}{N}[\d(\!r\!-\!y\!)\!
-\!\d(\!p\!-\!y\!)][\d(\!r\!-\!x\!)\!+\!\d(\!p\!-\!x\!)]
\tr S\dgg_uS_v\t^c\}\tr S\dgg_pS_r\t^a\t^b\}\d_{pr;uv}\Sigma \nn \\
& & \left.+\frac{1}{N^2}[\d(\!r\!-\!y\!)\!
-\!\d(\!p\!-\!y\!)][\d(\!t\!-\!x\!)\!-\!\d(\!q\!-\!x\!)]  
\tr S\dgg_uS_v\t^c\}\tr S\dgg_pS_r\t^b\}\tr S\dgg_qS_t\t^a\}\d_{qt;pr;uv}
\Sigma\right\},\nn 
\eea 

\bea
\label{lll}
f^{abc}&&\!\!\!\!J_L^a(x)J_L^b(y)J_L^c(w)\Sigma[s]= 
-\frac{1}{N}f^{abc}[\d(\!v\!-\!w\!)\!-\!\d(\!u\!-\!w\!)]\Bigl. 
\Bigr\{ \hspace{1cm} \\
 & &\qquad[\d(\!v\!-\!y\!)\!+\!\d(\!u\!-\!y\!)]
[\d(\!v\!-\!x\!)\!-\!\d(\!u\!-\!x\!)]
\tr S\dgg_u\t^c\t^b\t^aS_v\}\d_{uv}\Sigma \nn \\
 & &+\frac{1}{N}[\d(\!v\!-\!y\!)\!+\!\d(\!u\!-\!y\!)]
[\d(\!r\!-\!x\!)\!-\!\d(\!p\!-\!x\!)]
\tr S\dgg_u\t^c\t^bS_v\}\tr S\dgg_p\t^aS_r\}\d_{pr;uv}\Sigma \nn\\
& & +\frac{1}{N}[\d(\!r\!-\!y\!)\!
-\!\d(\!p\!-\!y\!)][\d(\!v\!-\!x\!)\!+\!\d(\!u\!-\!x\!)]
\tr S\dgg_u\t^c\t^aS_v\} \tr S\dgg_p\t^bS_r\}\d_{pr;uv}\Sigma \nn \\
& & +\frac{1}{N}[\d(\!r\!-\!y\!)\!
-\!\d(\!p\!-\!y\!)][\d(\!r\!-\!x\!)\!+\!\d(\!p\!-\!x\!)]
\tr S\dgg_u\t^cS_v\}\tr S\dgg_p\t^b\t^aS_r \}\d_{pr;uv}\Sigma \nn \\
 & & \left.+\frac{1}{N^2}[\d(\!r\!-\!y\!)\!
-\!\d(\!p\!-\!y\!)][\d(\!t\!-\!x\!)\!-\!\d(\!q\!-\!x\!)]  
\tr S\dgg_u\t^cS_v\}\tr S\dgg_p\t^bS_r\}\tr S\dgg_q\t^aS_t\}\d_{qt;pr;uv}
\Sigma \right\} , \nn
\eea

\bea
\label{lrr}
f^{acd}&&\!\!\!\!J_L^b(x)J_R^c(y)J_R^d(w)\Sigma[s]= 
-\frac{1}{N}f^{acd}[\d(\!v\!-\!w\!)\!-\!\d(\!u\!-\!w\!)]\Bigl. 
\Bigr\{ \hspace{1cm} \\
 & &\qquad[\d(\!v\!-\!y\!)\!+\!\d(\!u\!-\!y\!)]
[\d(\!v\!-\!x\!)\!-\!\d(\!u\!-\!x\!)]
\tr S\dgg_u\t^bS_v\t^c\t^d\}\d_{uv}\Sigma \nn \\
 & &+\frac{1}{N}[\d(\!v\!-\!y\!)\!+\!\d(\!u\!-\!y\!)]
[\d(\!r\!-\!x\!)\!-\!\d(\!p\!-\!x\!)]
\tr S\dgg_uS_v\t^c\t^d\}\tr S\dgg_p\t^bS_r\}\d_{pr;uv}\Sigma \nn\\
& & +\frac{1}{N}[\d(\!r\!-\!y\!)\!
-\!\d(\!p\!-\!y\!)][\d(\!v\!-\!x\!)\!-\!\d(\!u\!-\!x\!)]
\tr S\dgg_u\t^b\S_v\t^d\} \tr S\dgg_pS_r\t^c\}\d_{pr;uv}\Sigma \nn \\
& & +\frac{1}{N}[\d(\!r\!-\!y\!)\!
-\!\d(\!p\!-\!y\!)][\d(\!r\!-\!x\!)\!-\!\d(\!p\!-\!x\!)]
\tr S\dgg_uS_v\t^d\}\tr S\dgg_p\t^bS_r\t^c \}\d_{pr;uv}\Sigma \nn \\
 & & \left.+\frac{1}{N^2}[\d(\!r\!-\!y\!)\!
-\!\d(\!p\!-\!y\!)][\d(\!t\!-\!x\!)\!-\!\d(\!q\!-\!x\!)]  
\tr S\dgg_uS_v\t^d\}\tr S\dgg_pS_r\t^c\}\tr S\dgg_q\t^bS_t\}\d_{qt;pr;uv}
\Sigma \right\} , \nn
\eea
and
\bea
\label{llr}
f^{bcd}&&\!\!\!\!J_L^c(w)J_L^d(y)J_R^a(x)\Sigma[s]= 
-\frac{1}{N}f^{bdc}[\d(\!v\!-\!x\!)\!-\!\d(\!u\!-\!x\!)]\Bigl. 
\Bigr\{ \hspace{1cm} \\
 & &\qquad[\d(\!v\!-\!w\!)\!-\!\d(\!u\!-\!w\!)]
[\d(\!v\!-\!y\!)\!+\!\d(\!u\!-\!y\!)]
\tr S\dgg_u\t^d\t^cS_v\t^a\}\d_{uv}\Sigma \nn \\
 & &+\frac{1}{N}[\d(\!v\!-\!w\!)\!-\!\d(\!u\!-\!w\!)]
[\d(\!r\!-\!y\!)\!-\!\d(\!p\!-\!y\!)]
\tr S\dgg_u\t^dS_v\t^a\}\tr S\dgg_p\t^cS_r\}\d_{pr;uv}\Sigma \nn\\
& & +\frac{1}{N}[\d(\!r\!-\!w\!)\!
-\!\d(\!p\!-\!w\!)][\d(\!v\!-\!y\!)\!-\!\d(\!u\!-\!y\!)]
\tr S\dgg_u\t^cS_v\t^a\} \tr S\dgg_p\t^dS_r\}\d_{pr;uv}\Sigma \nn \\
& & +\frac{1}{N}[\d(\!r\!-\!w\!)\!
-\!\d(\!p\!-\!w\!)][\d(\!r\!-\!y\!)\!+\!\d(\!p\!-\!y\!)]
\tr S\dgg_uS_v\t^a\}\tr S\dgg_p\t^d\t^cS_r \}\d_{pr;uv}\Sigma \nn \\
 & & \left.+\frac{1}{N^2}[\d(\!r\!-\!w\!)\!
-\!\d(\!p\!-\!w\!)][\d(\!t\!-\!y\!)\!-\!\d(\!q\!-\!y\!)]  
\tr S\dgg_uS_v\t^a\}\tr S\dgg_p\t^dS_r\}\tr S\dgg_q\t^cS_t\}\d_{qt;pr;uv}
\Sigma \right\} . \nn
\eea

\subsection{Diagrammatic interpretation}
\label{diag}

Despite the fact that the expressions derived in the previous section 
appear complicated, they 
allow for a very clear physical interpretation in terms of diagrams according to
the following rules: the action of the left (right) rotation operator,
$J_{L(R)}^a$, on the 
projectile scattering matrix brings in a new dipole, along with the the 
corresponding $1/N$ suppression factor, 
which emits a new gluon of color $a$ before (or after) 
the interaction with the target. Such emission may happen either from the
quark line 
or from the antiquark. In the former case a, relative minus sign is picked up.
Subsequent actions of the rotation operators may act either on $\Sigma[s]$,
bringing new dipoles into the diagram, or on the preexisting dipole, which 
emits a new gluon. These rules are sketched in \fig{f1}.
\smallskip
\FIGURE{\epsfig{file=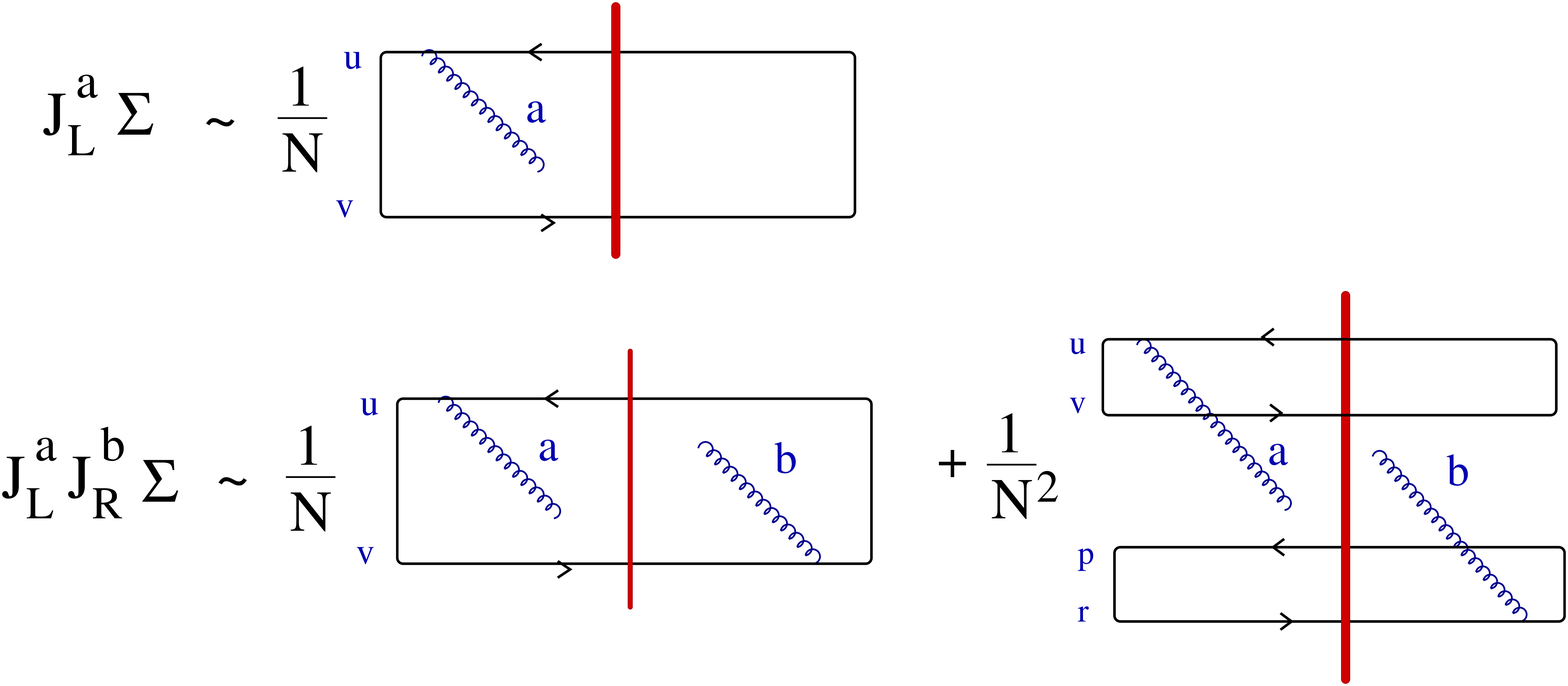,height=6cm,width=12cm} 
        \caption{Diagrammatic representation of $J_L^a\Sigma$ (top) and 
         $J_L^aJ_R^b\Sigma$ (bottom). The target is represented by the
         vertical thick line.}
        \label{f1}}

In order to fully describe the diagrams in terms of fundamental 
constituents, quark and antiquark lines, we
make use of the Fierz identity:
\beq
(t^a)_{ij}(t^b)_{kl}=\frac{1}{2}(\d_{il}\d_{jk}-\frac{1}{N}\d_{ij}\d_{kl}),
\label{fierz}
\eeq 
which can be translated into diagrammatic language substituting the 
gluon lines by quark-antiquark lines as indicated in \fig{f2}.
\smallskip
\FIGURE{\epsfig{file=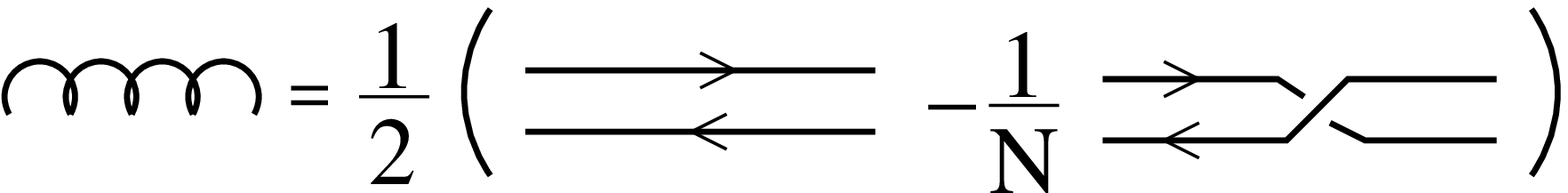,width=9cm} 
        \caption{Pictorial representation of the Fierz identity \eq{fierz}.}
        \label{f2}}

The propagation of each quark (antiquark) line through the
target at transverse coordinate $x$ is accounted for the scattering matrix, 
$S_F(x)$ ($S_F\dgg(x)$). Finally, the trace over closed fermion lines has to
be taken.
Following these rules, it is straightforward to rederive the results in
Eqs. (\ref{rrr})-(\ref{llr}).
As an example, we show in \fig{f3} the diagrams corresponding to the 
action of $J_LJ_LJ_L\Sigma[s]$, \eq{lll}, and $J_LJ_LJ_R\Sigma[s]$, \eq{llr},
in which, for the sake of simplicity, we have kept the gluon lines in the
adjoint representation without using the Fierz identity.

\smallskip
\FIGURE{\epsfig{file=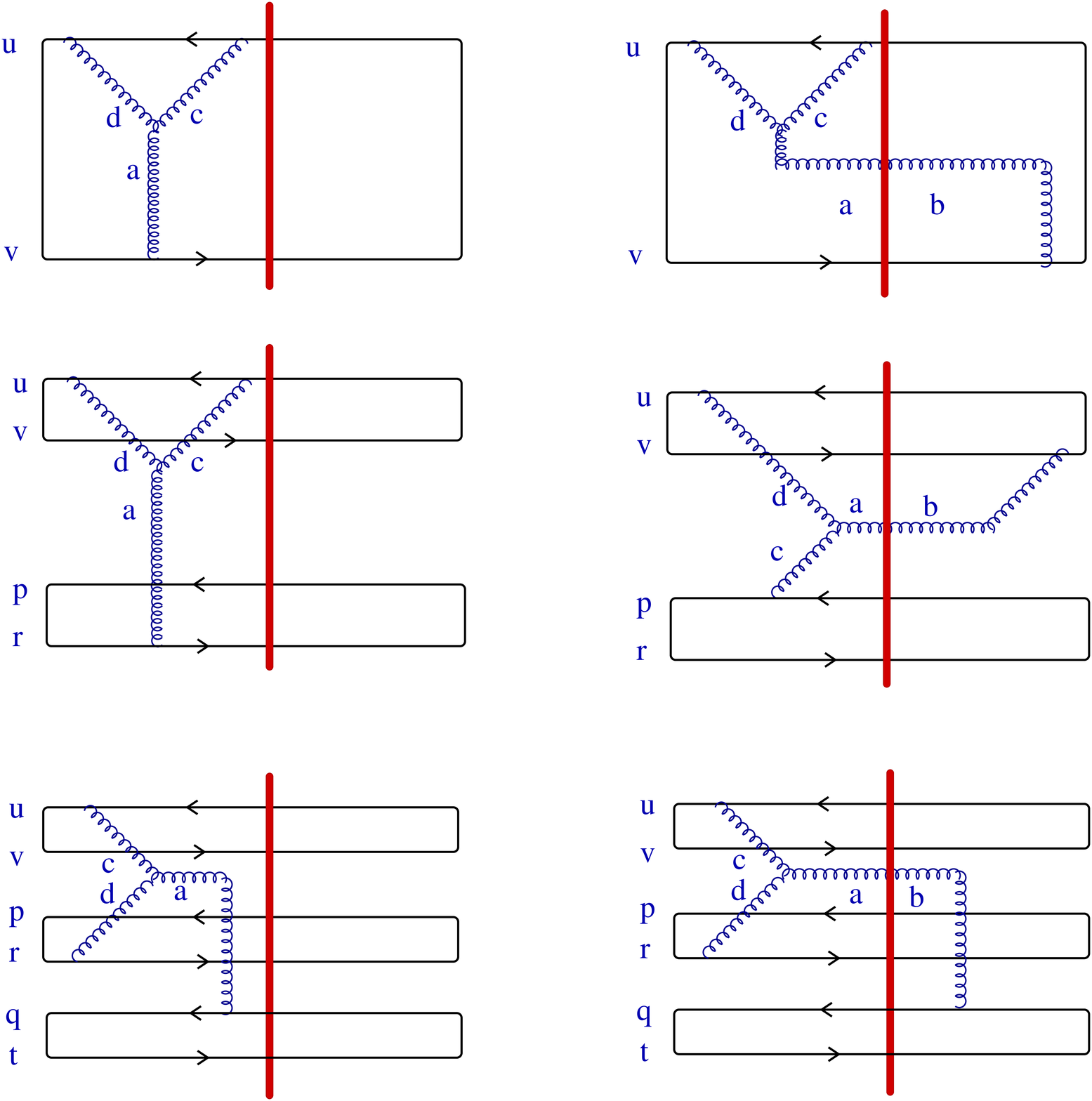,height=17cm,width=13cm} 
        \caption{Diagrams corresponding to the LLL (plots on the left) and LLR
(plots on the right), see the accompanying text. The
diagrams include one ($\propto 1/N$, top), two ($\propto 1/N^2$,
middle) and three  ($\propto 1/N^3$, bottom) active
dipoles. The thick vertical line represents the target.}
        \label{f3}}

\subsection{Symmetry considerations}
\label{symmetry}

Before proceeding further it is convenient to divide our final result 
into three pieces according to their leading power of $1/N$ as given by the
number of dipoles appearing in the relevant diagram, see the previous
subsection:
\beq
\chi^{2)}\Sigma[s]=\left(\chi^{2)}_{1/N}+\chi^{2)}_{1/N^2}+\chi^{2)}_{1/N^3}
\right)\Sigma[s].
\eeq
To complete the calculation we still have to plug in the results from 
Eqs. (\ref{rrr})-(\ref{llr}). into \eq{ker}, perform the contraction of the
color indices and integrate over  $x,y,w$.
Further, the cross terms LRR, \eq{lrr}, and LLR, \eq{llr}, must be contracted with the scattering matrix 
for a single gluon, $S_{A}^{ab}$, which can be rewritten in terms of
matrices in the fundamental representation by means of the Fierz identity:
\beq
S_{A}^{ab}(z)=2\,\tr S_F\dgg(z)\t^bS_F(z)\t^a\}.
\eeq  
The $SU(N)$ algebra can be worked out using the 
relations listed in the Appendix. 

After all this is done, we note that the contribution to the kernel
coming from the 
pieces of order $1/N$ in Eqs. (\ref{rrr}) and (\ref{lll}) cancel each other
out.
An analogous cancellation happens between the leading $1/N$ contributions from
Eqs. (\ref{lrr}) and (\ref{llr}). Therefore
\beq
\chi^{2)}_{1/N}\Sigma[s]=0.
\label{n1}
\eeq 

It is clear from the diagrammatic rules derived in \ref{diag} that this
cancellation corresponds to the diagrams in which all the 
emissions and absorptions of gluons happen in a single dipole, while the other
dipoles in the wave function remain as spectators. More precisely, the top
diagrams in \fig{f3} cancel against their respective complex conjugate
diagrams (not shown in the figure).

An analogous cancellation occurs for the $1/N^2$ terms in 
Eqs. (\ref{rrr}) and (\ref{lll}). Diagrammatically, the middle-left diagram in
\fig{f3} cancels against its complex conjugate. Therefore, the leading $1/N^2$
contribution is given by the $1/N^2$ terms in Eqs. (\ref{lrr}) and
(\ref{llr}) (middle-right diagram in \fig{f3} plus its complex conjugate). 
We get
\bem
\chi^{2)}_{1/N^2}\Sigma[s]=
\frac{i}{(2\pi)^2}\frac{1}{N^2}\Biggl\{\int d^2z\\
-N[\partial_i(R\!-\!P)][U\parf V][\tr S\dgg_uS_vS\dgg_zS_rS\dgg_pS_z\}
-\frac{1}{N}\tr S\dgg_uS_v\}\tr S\dgg_pS_r\}]\\
-\frac{1}{2}\,[\partial_i(V\!-\!U)][(R\!-\!P)\parf(V\!-\!U)]
[\tr S\dgg_zS_vS\dgg_pS_r\}\tr S\dgg_uS_z\}
-\tr S\dgg_uS_zS\dgg_pS_r\}\tr S\dgg_zS_v\}]\\
+\frac{1}{2}\,[\partial_i(R\!-\!P)][(R\!-\!P)\parf(V\!-\!U)]
[\tr S\dgg_zS_rS\dgg_uS_v\}\tr S\dgg_pS_z\}-
\tr S\dgg_pS_zS\dgg_uS_v\}\tr S\dgg_zS_r\}]\\
-\frac{1}{2}\,[\partial_i(V\!-\!U)][(R\!-\!P)\parf(V\!-\!U)]
[\tr S_zS\dgg_uS_rS\dgg_p\}\tr S\dgg_zS_v\}-
\tr S\dgg_pS_vS\dgg_zS_r\}\tr S\dgg_uS_z\}]\\
+\frac{1}{2}\,[\partial_i(V\!-\!U)][(V\!-\!U)\parf(R\!-\!P)]
[\tr S_zS\dgg_uS_rS\dgg_p\}\tr S\dgg_zS_v\}-
\tr S\dgg_pS_vS\dgg_zS_r\}\tr S\dgg_uS_z\}]\\
+{N}[\partial_i(V\!-\!U)][P\parf R][\tr S\dgg_uS_vS\dgg_zS_rS\dgg_pS_z\}-
\frac{1}{N}\tr S\dgg_uS_v\}\tr S\dgg_pS_r\}]\Biggr\}\,\frac{\d^2 \Sigma[s]}{\d
  s(p,r)\d s(u,v)}\,.
\label{n21}
\end{multline} 

This result can be further simplified by noting that any wave function or
weight functional of a gluonic/dipole configuration has to be completely 
symmetric under the exchange of any number of gluons/dipoles, such that 
the exchange $(u,v)\leftrightarrow (p,r)$ leaves
the action of the functional derivative $\delta_{uv;pr}\Sigma$
unchanged. Under such an exchange, many terms in \eq{n21} cancel each other,
yielding:  

\bea
& &\chi^{2)}_{1/N^2}\Sigma[s]=-\frac{i}{(2\pi)^2} \,\frac{1}{N^2}
\int d^2z\,\Big\{\nn
\\
& &[\partial_i(R\!-\!P) (U\!\parf\! V)-
\partial(V\!-\!U) (P\!\parf\!R)] [
N\,\tr S_u^\dagger S_v S_z^\dagger S_rS_p^\dagger S_z \}]\nn
\\
& &
+\partial_i(V\!-\!U)
(R\!-\!P)\parf (V\!-\!U)
\left[ \tr S_z^\dagger S_v S_p^\dagger S_r \} \tr  S_u^\dagger
S_z\}-\tr S_u^\dagger S_z S_p^\dagger S_r\} \tr  S_z^\dagger S_v\}\right.
\nn \\ 
& &\left.  
 +\tr S_p^\dagger S_zS_u^\dagger S_r\}\tr  S_z^\dagger S_v\}
-\tr  S_p^\dagger S_vS_z^\dagger S_r\}\tr  S_u^\dagger
S_z\}\right]\Big{\}}\,\frac{\d^2\Sigma[s]}{\d s(p,r)\d s(u,v)}\,\,.
\label{n2}
\eea   

Calculating the $1/N^3$ terms in an analogous way we get:          
\bea
& &\chi^{2)}_{1/N^3}\Sigma[s]=-\frac{i}{2}\,\frac{1}{(2\pi)^2} \,\frac{1}{N^3}
\int d^2z \left[ \{\partial_i(R\!-\!P)\}
\{(T\!-\!Q)\parf (V\!-\!U)\}\right]
\nn \\
& &\times \left[\tr S^\dagger_uS_vS^\dagger_qS_tS^\dagger_pS_r\}
+\tr S^\dagger_uS_tS^\dagger_qS_rS^\dagger_pS_v\} +
\tr S^\dagger_zS_rS^\dagger_pS_zS^\dagger_uS_vS^\dagger_qS_t\}
\right.
\nn \\
& &\left. 
+\tr S^\dagger_pS_rS^\dagger_zS_vS^\dagger_uS_t S^\dagger_qS_z\}\right]
\frac{\d^3\Sigma[s]}{\d s(q,t)\d s(p,r)\d s(u,v)}\,\,.
\label{n3}
\eea

Eqs. (\ref{n1}), (\ref{n2}) and (\ref{n3}) are the central result of this
paper. 
One of their main features is the fact that the action of the kernel on a
dipole-like projectile cannot be entirely recast in terms of dipole degrees of 
freedom. On the contrary, the evolution generates a complicated color 
structure consisting in the mixing of dipoles, quadrupoles, sextupoles 
and octupoles as given by these equations, which can only be expressed
partially as dipole degrees of freedom by increasing the power in $1/N$.
Consequently, these
corrections to the leading JIMWLK kernel bring no corrections to the mean
field BK equation.

\vskip -1.5cm
\section{Conclusions}
\label{conclu}
In this work we have calculated the $\mathcal{O}(\alpha_s^2)$ corrections to
the JIMWLK kernel that arise from the $\mathcal{O}(g^3)$
solutions to the classical
Yang-Mills equations. These are the first corrections in the charge density of
the projectile to JIMWLK evolution, and  therefore partially account for the
coherence effects in the
projectile gluon emission which drives small-$x$ evolution. In the context of
the present discussions on the quest for evolution equations for scattering of
a dense projectile which would contain the so-called pomeron loops, our result
accounts for the part of these corrections not arising from the
non-commutativity of the target fields. Thus, they are restricted to the
dilute-dense scattering situation. The
same systematic technique could be used to improve the results,
although admittedly
the computation of even higher orders looks technically challenging.

Our main result is the cancellation of the leading $1/N$ contributions,
together with Eqs. (\ref{n2})-(\ref{n3}). We provide a
diagrammatic interpretation of these results.
These corrections are subleading in $1/N$ and exhibit a complicated color
structure. They do not, therefore,  provide any correction to the BK equation
but, rather,  add new terms to the Balitsky hierarchy.

\section*{Acknowledgments}
It is a pleasure to thank Yuri Kovchegov, Alex Kovner, Misha Lublinsky and 
Heribert Weigert for most useful discussions and a critical reading of this
manuscript, and Genya Levin for enlightening comments.
JLA and JGM thank the Departamento de 
F\'{\i}sica de Part\'{\i}culas of Universidade de Santiago de Compostela, and
JLA and NA thank CENTRA/IST for warm hospitality during stays
when part of this work was done.

\appendix
\label{app}
\section{Some color algebra}

In this appendix we list some relations needed to derive the results in
section \ref{dipole}. Using the basic $SU(N)$ relations
\begin{align}
& f^{ade}f^{bef}f^{cfd}=\frac{N}{2}f^{abc}\,,\\
& d^{ade}d^{bef}f^{cfd}=\left(\frac{2}{N}-\frac{N}{
   2}\right)f^{abc}\,,\\
& d^{ade}f^{bef}f^{cfd}=-\frac{N}{2}\,d^{abc}\,,\\ 
& d^{ade}d^{bef}d^{cfd}={N^2-12\over2N}\,d^{abc}\,,\\
& \tr t^at^bt^ct^d\}=\frac{1}{4N}\,\d^{ab}\d^{cd}+\frac{1}{8}\,g^{abe}g^{cde}\,,
\end{align}
where we have introduced the notation:
\beq
g^{abc}\equiv4\,\tr t^at^bt^c\}=(d^{abc}+if^{abc});
\qquad\bar{g}^{abc}\equiv(d^{abc}-if^{abc})\,.
\eeq

\noi Since the generators of $SU(N)$, $t^a$, along with the unit
matrix form a basis of the matrices in the fundamental representation, we
expand any arbitrary matrices, $X,Y\dots$ in the following way
\beq
X=x_0+x_1^a\t^a,\quad Y=y_0+y_1^a\t^a,\dots
\eeq

\noi With the help of the relations listed above, we get

\begin{align}
& f^{abc}\,\tr XYt^at^bt^c\}=\frac{i}{4}(N^2-1)\tr XY\},\\
& f^{abc}\,\tr Xt^aYt^bt^c\}=\frac{iN}{4}\left(\tr X\}\tr Y\}-\frac{1}{N}
\tr XY\}  \right), \\
& f^{abc}\,\tr Xt^aYt^b\}\tr Zt^c\}=\frac{i}{4}\left[\tr XZ\}\tr Y\}-\tr YZ\}
  \tr X\}\right],\\
& f^{abc}\,\tr XYt^at^b\}\tr Zt^c\}=\frac{iN}{4}\left(\tr XYZ\}-\frac{1}{N}\tr XY\}
  \tr Z\} \right),\\
& f^{abc}\,\tr Xt^a\}\tr Yt^b\}\tr Zt^c\}=\frac{i}{4}\left[\tr XZY\}-\tr XYZ\}
  \right].
\end{align}

\noi These are the relations required to perform the
contraction of the color indices in Eqs. (\ref{rrr})-(\ref{llr}).\


\providecommand{\href}[2]{#2}\begingroup\raggedright\endgroup

\end{document}